# Field - Effect Switching in Nano-Graphite Films

Sergey G. Lebedev

*Department of Experimental Physics,* Institute for Nuclear Research of Russian Academy of Sciences, Moscow, 117312, Russian Federation

The effect of electrical resistivity switching in nano – graphite films is described. In difference with cases published elsewhere the switching in nano – graphite films occurs from stable high conductive to metastable low conductive state. Critical current of switching varies in the range *0.01-0.5A* and can be increased up to values appropriate for using of nano - graphite samples in power grids as contact-less current limiters and circuit breakers. The possible mechanisms of switching phenomenon in nano – graphite films are discussed.

Physical and electromagnetic properties of the carbon and its derivatives are the subjects of constant interest and extensive studies for many years. Attention has been paid to the switching behavior in the carbon graphite like materials. K. Antonowicz more than *30* years ago investigated the conductive properties of the glassy carbon [1] and its evaporated deposits [2] and has found out the effect of jump of conductivity up to three orders of magnitude. The change of conductivity was reversible, and the relaxation time made some days.

B.Z.Jang and L.Zhao also studied the switching behavior of carbonized materials [3]. They study the partially carbonized polyacrylnitride fibbers, which were observed to undergo a resistivity change between *2* and *4* orders of magnitude at a transition temperature typically in the range of *98°C* to *200°C*. The current-voltage curves exhibited an initial supercurrent-like increase, followed by a rapid drop to a high resistance state, and then a rise in current again at a later stage.

H.A.Goldberg et all have obtained the *US* patent on the carbon switching device [4]. This device was made on the base of partially pyrolyzing polymer material by means of heating of the material to between *500°C* and *800°C*. Electrodes are connected with the material at two different locations to produce an electrically active element therebetween. Devices made according to the teachings of the disclosure exhibit the negative resistance in a part of their voltage-current characteristic and run as the bi-directional electronic switches.

The author of this paper spent more than twenty years for study of the thin carbon films having in mind their application first as the strippers for the charge particle beams. Later on the attention has been paid to the electromagnetic properties of thin carbon films. Some anomalies in the electromagnetics of the films formed by sputtering of spectroscopic pure graphite in electrical arc discharge and also by chemical vapor deposition (CVD) methods have been found [5-8]. This was the field effect electrical resistivity jump, which can be used for current limiting in electric circuits and smart chips, the *rf*-to-*dc* conversion and the optical radiation emitted after switching off, which can utilized in various optoelectronic devices.

In this Letter we report about switching behavior observed in NG films, which can be used potentially as current limiters in smart grids.

The nano-graphite films using in these experiments were obtained by means of CVD method on the quartz substrate. CVD process has been take place in thermal activated hydrogen-acetylene gas mixture at the temperature of *900°C*. As a result the *1* micrometer thick NG film has been produced. Figure 1 represent typical scanning electron microscopy (SEM) image of NG film. As can be seen at highest magnification of *20000* the surface of deposited film look likes structureless which means the film is amorphous. This result is consistent with the representation [9] about NG films as a composite of small graphite-like $sp^2$-bonding granules embedded in the matrix of amorphous carbon. This conclusion is supported by in-situ Raman spectroscopy.

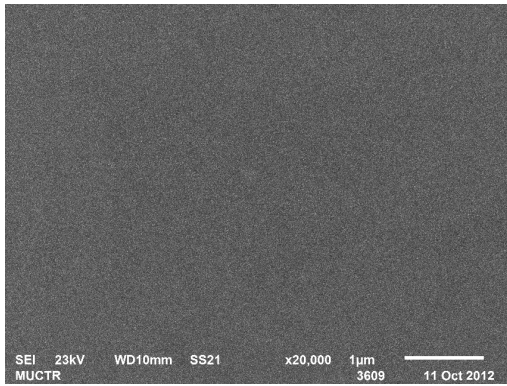

Figure1. SEM image of typical nano-graphite film with the magnification of *20000*.

The typical Raman spectrum of NG film is shown in Fig.2. In this spectrum the Raman D-band at *1360 cm$^{-1}$* and G-band in the vicinity of *1600 cm$^{-1}$* are specific for different forms of disordered graphite. The height equality of D and G peaks corresponds to graphite crystalline size of the order *20-30 A* [10].

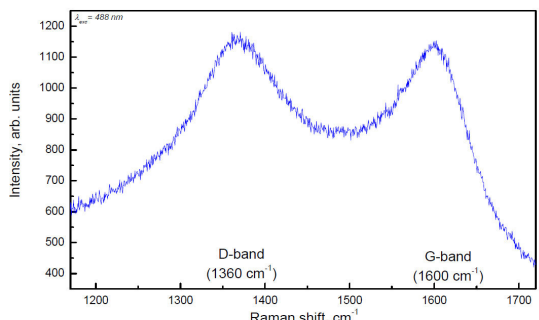

Figure 2. Raman spectrum of typical nano-graphite film. G and D – bands attributed to different forms of disordered graphite are clearly seen.

For the conductive measurements we used the samples of *1-2* cm$^2$ square with copper current leads arranged in planar geometry and connected to film surface by means of a silver paste. The *I-V* and switching characteristics of these samples were studied using both LabVIEW and manual controlling current source.

The current-voltage and electrical switching characteristics of NG film samples at room temperature are shown in Fig.3. In the "ON" state with the increasing voltage the current through the sample initially increase linearly (Fig.3). Near the critical current the samples show nonlinearity. At the critical current the samples exhibited jump into "OFF" state with the resistivity four to five order of magnitude higher of that into the "ON" state. Such a resistivity jump look like to metal – insulator transition. The value of critical current varies in the range *0.01-0.5A*. This value is limited by difference in thermal expansion of NG film and substrate. When the applied voltage decreases to zero the sample reverts into the "ON" state and the switcher can be used repeatedly many times. As can be seen from Fig. 3 the change of resistivity up to four to five orders of magnitude occurred as a result of switching.

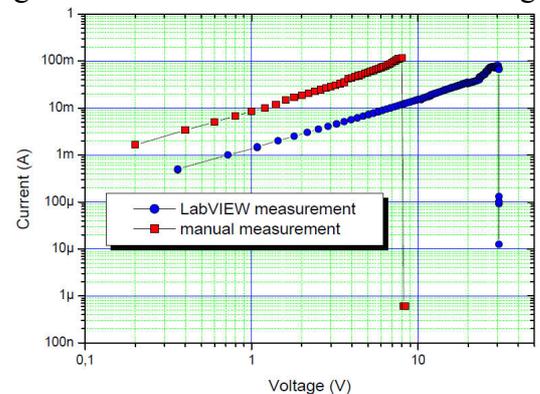

Figure 3. Room temperature *I-V* characteristics of 1 micrometer layer of NG film on quartz substrate.

The switching rate measurement has been made by oscilloscope on load resistance. The switching process is shown on Fig.4. As can be seen the duration of switching process is about *200* microseconds. This value varies up to *50%* from sample to sample.

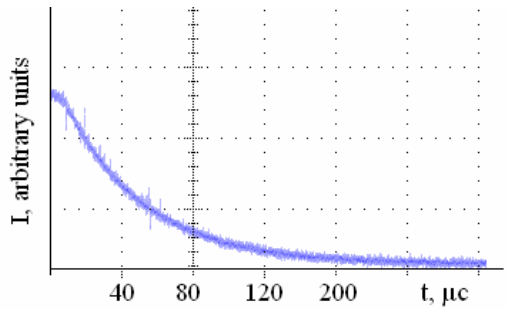

Figure 4. Switching rate of NG film sample.

Such switching behavior is drastically different from other switchers described elsewhere [11-13]. The essential of all of these devices is switching from the low conducting (LC) to high conducting (HC) state. For explanation of such kind of switching numerous hypotheses has been proposed: thermal and double injection mechanism [14], exciton's mechanism [11] and bistability caused by the overheating of electrons [15-18]. On our opinion none of these hypotheses can explain our results because in our case the switching occurs from HC to LC state that is reversely to processes mentioned above.

Thermal mechanism falls to explain our results due to fast switching with "electronic" times of *100-200* microseconds (see Fig. 4). Also we observed optical radiation emitted few milliseconds after switching off [7]. It is believed this process is related with thermo cooling of sample after switching.

Double injection mechanism suggests the existence of trap centers in the structure. Saturation of these centers with injected curriers gives rise to jump into LC state. However it is not possible to reverse this process to produce LC to HC transition.

As for the hypothesis of hot electrons the steep change of resistivity on few orders of magnitude should be located in near vicinity of working temperature. In the case described in [15-16] this is the superconductor – insulator transition near the critical temperature. In Ref. [3] the evidence of steep metal – semiconductor transition between HC and LC around a temperature of *425$^o$K* is presented. In our case there are no some definite proofs for existence of such kind transition around room temperature.

The superconducting hypothesis [19] can explain the results obtained in [1-3] as well as the switching behavior of NG films and other our results [5-8]. First of all the superconducting behavior should be related with steep transition into zero resistive (or high conductive) state. Though the metal – insulator transitions in graphite [20-22] and disordered carbon [23-24] are well known, however as it has been mentioned before so far in NG films we have not find steep transition into insulating state above room temperature. But the main disadvantages of superconducting hypothesis are the absence of zero resistance state and Meissner effect in all experiments pointing above. Both contradictions can be explained by movement of magnetic flux quanta in the type II superconductor with sufficiently small lover critical magnetic field. But how we can prove the superconductivity in this case? Is it possible to distinguish between semiconductor and type II superconductor? One of the possibilities to shed some light on the above questions may be the measurement of thermal balance of sample. During the movement of vortices the light emission occurs which gives rise to additional cooling of sample. Of course the other methods also should be used such as measurements in high magnetic field and temperatures in trying to find the upper critical magnetic field and superconductor – insulator transition respectively. Also *I-V* measurements at very small currents might be useful in trying to find supercurrent behavior.

So the switching mechanism in NG film so far is a puzzle. Nevertheless it could be expected that due to its high critical current at room temperature NG film is likely a kind of material with prospective applications in smart grids as a current limiters and circuit breakers.

In conclusion, switching effects from HC to LC in NG films are described. It is shown the switching mechanism cannot be explained by existing models. The critical current of switching varies in the range *0.01-0.5A*. It is believed this current can be increased up to the values appropriate for using of NG samples as contactless current limiters and circuit breakers for power grids.

This work was done with the support of the Russian Foundation for Basic Research (RFBR) under Grant No. 05-08-17909-a. The author wish to thank A. Saveliev for producing SEM and Raman measurements.

[1] K. Antonowicz, L. Cacha, J. Turlo, Carbon **11**, 1 (1973).
[2] K. Antonowicz, A. Jesmansowicz, and J. Wieczorek, Carbon **10**, 81 (1972).
[3] B. Z. Jang, and L. R. Zhao, Journal of Material Research **10**, 2449 (1995).
[4] H. A. Goldberg, I. L. Kalnin, and C. C. Williams, US Patent 4,642,664, (1987).
[5] S. G. Lebedev, and S. V. Topalov, Bulletin of Lebedev's Physical Institute **N11- 12**, 14 (1994).
[6] S. G. Lebedev, Nuclear Instruments and Methods in Physics Research **A521**, 22 (2004).
[7] S. G. Lebedev, V. E. Yants, and A. S. Lebedev, Nuclear Instruments and Methods in Physics Research **A590**, 227 (2008).
[8] S. G. Lebedev, International Review of Physics (IREPHY) **2**, 312 (2008).
[9] G. P. Lopinski, V. I. Merkulov, and J. S. Lannin, Physical Review Letters, **80**, 4241 (1998).
[10] P. K. Chu, L. Li, Materials Chemistry and Physics **96,** 253 (2006).
[11] L. R. Zhao, B. Z. Jang, Journal of Material Sciences Letters **15,** 99 (1996).
[12] S. R. Ovshinsky, Physical Review Letters, **21**, 1450 (1968).
[13] S. S. K. Titus, R. Chatterjee, S. Asokan, and A. Kumar, Physical Review **B48,** 14650 (1993).
[14] N. F. Mott, Philosophical Magazine **24,** 911 (1971).
[15] B. L. Altshuler, V. E. Kravtsov, I. V. Lerner, and I. L. Aleiner, Physical Review Letters **102,** 176803 (2009).
[16] M. Ovadia, B. Sacepe, and D. Shahar, Physical Review Letters **102**, 176802 (2009).
[17] F. Ladieu, M. Sanquer, and J. P. Bouchaud, Physical Review **B53**, 973(1996).
[18] A. Zherebov, A. Lachinov, J. Genoe, and A. Tameev, Applied Physics Letters **92**, 193302(2008).
[19] K. Antonowicz, Nature **247,** 358-360 (1974).
[20] J. C. Gonzalez, M. Munoz, N. Garcıa, J. Barzola-Quiquia, D. Spoddig, K. Schindler, and P. Esquinazi, Physical Review Letters **99,** 216601 (2007).
[21] G. Timp, P. D. Dresselhaus, T. C. Chieu, G. Dresselhaus, and Y. Iye, Physical Review **B28**, 7393(1983).
[22] Z.M. Wang, Q.Y. Xu, G.Ni, and Y.W. Du, Physics Letters **A314**, 328(2003).
[23] K. Kuriyama, and M. S. Dresselhaus, Journal of Material Research 7, 940(1992).
[24] A. W. P. Fung, Z. H. Wang, M. S. Dresselhaus, G. Dresselhaus, R. W. Pekala, and M. Endo, Physical Review **B49**, 17325(1994).